\documentclass{article}

\usepackage{amssymb}
\usepackage{epsfig}
\usepackage{color}
\usepackage{amsfonts}
\usepackage{amsmath}

\reversemarginpar
\newcommand{\spacer}{\rule[0cm]{0cm}{0cm}}

\newcounter{statementnumber}
\renewcommand{\thestatementnumber}{\arabic{section}.\arabic{statementnumber}}

\newenvironment{thm}[2]{\refstepcounter{statementnumber} \label{#2}
\par \noindent {\bf #1~\thestatementnumber.}  
\begin{em}}{\end{em} \par}

\newenvironment{proof}{\par\noindent\textsc{Proof.}}{\nopagebreak\spacer\hfill $\square$}

\newcounter{figurecount}
\def\mb{$\begin{displaystyle}}
\def\me{\end{displaystyle}$\ }

\def\rank{\mbox{\rm rank}}
\newcommand{\ket}[1]{\left| #1 \right\rangle}
\newcommand{\bra}[1]{\left\langle #1 \right|}

\newcommand{\twotwo}[4]{\left[ \begin{array}{cc} #1 & #2 \\
                        #3 & #4 \end{array} \right]}

\newcommand{\R}{{\mathbb R}}

\newcommand{\Proj}{{\mathbb P}}
\newcommand{\C}{{\mathbb C}}

\begin{document}

\thispagestyle{empty}

\begin{center}
{\Large \bf Classification of $n$-qubit states\\
%\vspace*{-.5ex}
with minimum orbit dimension}\\

\medskip

David W. Lyons\\
lyons@lvc.edu\\
Mathematical Sciences\\
Lebanon Valley College\\

\medskip
Scott N. Walck\\
walck@lvc.edu\\
Department of Physics\\
Lebanon Valley College\\
\medskip

revised:  19 January 2006\\
\end{center}

%\vspace*{.2in}

%\pagenumbering{roman}
\thispagestyle{empty}

%\tableofcontents

%\newpage
%\pagenumbering{arabic}
%\setcounter{page}{1}

% \setcounter{section}{-1}

\begin{center}
\begin{minipage}{4in}
{\bf Abstract}. The group of local unitary transformations acts on the
space of $n$-qubit pure states, decomposing it into orbits.  In a
previous paper we proved that a product of singlet states (together with
an unentangled qubit for a system with an odd number of qubits) achieves
the smallest possible orbit dimension, equal to $3 n / 2$ for $n$ even
and $(3 n + 1)/2$ for $n$ odd, where $n$ is the number of qubits.  In
this paper we show that any state with minimum orbit dimension must be
of this form, and furthermore, such states are classified up to local
unitary equivalence by the sets of pairs of qubits entangled in singlets.
\end{minipage}
\end{center}

\setcounter{statementnumber}{0}
\section{Introduction}

Quantum entanglement has been identified as an important resource for
quantum computing, quantum communication, and other applications
\cite{nielsenchuang,gudder03}.  A fundamental theoretical problem is to
understand the types of entanglement that composite quantum systems can
achieve.

Defining entanglement types as equivalence classes of quantum states
under local unitary (LU) equivalence is perhaps the most natural way to
proceed in classifying entanglement \cite{linden98,linden99}.  The group
of LU transformations acts on the space of quantum states, partitioning
it into LU orbits.  Each orbit is a collection of locally equivalent
quantum states that forms a differentiable manifold with a certain
dimension.

The classification of entanglement types has turned out to be a
difficult problem.  Most of the progress in understanding multipartite
entanglement has been for systems of only two or three
qubits \cite{acin00,acin01,carteret00a}.  Few results exist concerning
the classification of $n$-qubit entanglement types for arbitrary $n$.

A promising approach to the difficult problem of characterizing
entanglement types is to break the problem into two parts.  First,
identify the possible dimensions of LU orbits in the state space.  Then,
identify the orbits that have each possible dimension.  In
\cite{lyonswalck}, the present authors addressed the first part of this
program by identifying the allowable range of orbit dimensions for
$n$-qubit quantum states to be $3n/2$ to $3n$ for even $n$ and
$(3n+1)/2$ to $3n$ for odd $n$.  The present paper begins the second
part of the program by completely identifying all orbits (entanglement
types) with minimum dimension.

In this paper, we identify all $n$-qubit entanglement types that have
minimum orbit dimension.  States that have the minimum orbit dimension
are, in some sense, the ``rarest'' quantum states.  We show that the
only quantum states to achieve minimum orbit dimension are tensor
products of singlet states (with a single unentangled qubit for $n$ odd)
and their LU equivalents.  This suggests a special role for the 2-qubit
singlet state in the theory of $n$-qubit quantum entanglement.

\setcounter{statementnumber}{0}
\section{Notation and previous results}

In this section we establish notation, some definitions, and state some
results from our previous paper~\cite{lyonswalck} needed for the present
paper.  For the convenience of the reader, we give a list (Appendix~A)
of equation and statement numbers in the present paper with their
matching numbers in~\cite{lyonswalck}.

Let $|0\rangle$, $|1\rangle$ denote the standard computational basis for
$\C^2$ and write $|i_{1}i_{2}\ldots i_n\rangle$ for
$|i_{1}\rangle\otimes|i_{2}\rangle\otimes\cdots\otimes|i_n\rangle$ in
$(\C^2)^{\otimes n}$.  For a multi-index $I=(i_{1}i_{2}\ldots i_n)$ with
$i_k = 0,1$ for $1\leq k\leq n$, we will write $|I\rangle$ to denote
$|i_{1}i_{2}\ldots i_n\rangle$.  Let $i_k^c$ denote the bit complement
$$i_k^c = \left\{\begin{array}{ll}
0 & \mbox{if $i_k = 1$}\\
1 & \mbox{if $i_k = 0$}
\end{array}\right.$$
and let $I_k$ denote the multi-index 
$$I_k:=(i_{1}i_{2}\ldots i_{k-1} i_k^c i_{k+1} \ldots i_n)$$ obtained from
$I$ by taking the complement of the $k$th bit for $1\leq k\leq n$.
Similarly, let $I_{kl}$ denote the multi-index 
$$I_{kl}:=(i_{1}i_{2}\ldots i_{k-1} i_k^c i_{k+1} \ldots i_{l-1} i_l^c
i_{l+1} \ldots i_n)$$ obtained from $I$ by taking the complement of the
$k$th and $l$th bits for $1\leq k<l\leq n$.

Let $H= (\C^2)^{\otimes n}$ denote the Hilbert space for a system of $n$
qubits and let $\Proj(H)$ denote the projectivization of $H$ which is
the state space of the system.  We take the local unitary group to be
$G=\mbox{\rm SU}(2)^n$.  Its Lie algebra $LG=\mbox{\rm su}(2)^n$ is the
space of $n$-tuples of traceless skew-Hermitian $2\times 2$ matrices.
We choose $\displaystyle A = i\sigma_z = \twotwo{i}{0}{0}{-i}$,
$\displaystyle B = i\sigma_y = \twotwo{0}{1}{-1}{0}$, and $\displaystyle
C = i\sigma_x = \twotwo{0}{i}{i}{0}$ as a basis for $\mbox{\rm su}(2)$, where
$\sigma_x,\sigma_y,\sigma_z$ are the Pauli spin matrices.  We define
elements $A_k,B_k,C_k$ of $LG$ for $1\leq k\leq n$ to have $A,B,C$,
respectively, in the $k$th coordinate and zero elsewhere.
\begin{eqnarray*}
A_k &=& (0,\ldots,0,\twotwo{i}{0}{0}{-i},0,\ldots,0)\\
B_k &=& (0,\ldots,0,\twotwo{0}{1}{-1}{0},0,\ldots,0)\\
C_k &=& (0,\ldots,0,\twotwo{0}{i}{i}{0},0,\ldots,0)
\end{eqnarray*}
Given a state vector $\ket{\psi}=\sum c_I\ket{I}$, we have the following.
\begin{eqnarray}
  A_k\ket{\psi} &=& \sum_{I} i(-1)^{i_k}c_{I} \ket{I}\label{akact}\\
  B_k\ket{\psi} &=& \sum_{I} (-1)^{i_k}c_{I_k} \ket{I}\label{bkact}\\
  C_k\ket{\psi} &=& \sum_{I} ic_{I_k} \ket{I}\label{ckact}
\end{eqnarray}
Given a state vector $\ket{\psi}\in H$, we define the $2^n\times(3n+1)$
complex matrix $M$ to be the $(3n+1)$-tuple of column vectors
\begin{equation}\label{colsofMident}
M=(A_1\ket{\psi},B_1\ket{\psi},C_1\ket{\psi},
\ldots,A_n\ket{\psi},B_n\ket{\psi},C_n\ket{\psi},-i\ket{\psi}).
\end{equation}
We view the matrix $M$ and also its column
vectors both as real and also as complex via the standard identification 
\begin{eqnarray}
\C^N &\leftrightarrow &\R^{2N} \nonumber \\
(z_1,z_2,\ldots,z_N) &\leftrightarrow &(a_1,b_1,a_2,b_2,\ldots,a_N,b_N) \label{cnidentr2n}
\end{eqnarray}
where $z_j = a_j + ib_j$ for $1\leq j\leq N$.
The real rank of $M$ gives the
dimension of the $G$ orbit of a state.
\begin{thm}{Proposition}{orbitdimasrank}
{Orbit dimension as rank of $M$:} Let $x\in \Proj(H)$ be a state,
let $\ket{\psi}$ be a Hilbert space representative for $x$, and let $M$ be the
associated matrix constructed from the coordinates of $\ket{\psi}$ as described
above.  Then we have
$$\rank_{\R} M = \dim {\cal O}_x + 1.$$
\end{thm}

We denote by ${\cal
  C}$ the set 
$${\cal C}=\{A_1\ket{\psi},B_1\ket{\psi},C_1\ket{\psi},
\ldots,A_n\ket{\psi},B_n\ket{\psi},C_n\ket{\psi},-i\ket{\psi}\}$$ of
columns of $M$, and for $1\leq k\leq n$ we define
the {\em triple $T_k\subseteq{\cal C}$} to be the set of vectors
\begin{equation}
T_k=\{A_k\ket{\psi},B_k\ket{\psi},C_k\ket{\psi}\}.
\end{equation}
For a subset ${\cal B}\subseteq {\cal C}$, we write $\langle {\cal
B}\rangle$ to denote the real span of the column vectors contained in
${\cal B}$.  For convenience we write $\langle
T_{i_1},T_{i_2},\ldots,T_{i_r}\rangle$ to denote the real span $\langle
T_{i_1}\cup T_{i_2}\cup \cdots\cup T_{i_r}\rangle$ of a set of triples.
We have the following facts about the dimensions of spans of sets of
triples.

\begin{thm}{Proposition}{triplesprop}
%{Triples span three dimensions:} 
Let $T_k=\{A_k\ket{\psi},B_k\ket{\psi},C_k\ket{\psi}\}$ be a triple of
columns of $M$.  The three vectors in the triple are orthogonal when
viewed as real vectors.
\end{thm}

\begin{thm}{Proposition}{mainpropgen}
{Main orthogonality proposition:} Suppose that
$$\dim\langle T_{j_1},T_{j_2},\ldots,T_{j_m}\rangle<3m$$ for some $1\leq
j_1<j_2<\cdots<j_m\leq n$.  Then there is a nonempty subset $K\subseteq
\{j_1,j_2,\ldots,j_m\}$ containing an even number of elements such that
there are two orthogonal vectors $\ket{\zeta_k},\ket{\eta_k}$ in
$\langle T_k \rangle$, both of which are orthogonal to $-i\ket{\psi}$,
$A_j\ket{\psi}$, $B_j\ket{\psi}$ and to $C_j\ket{\psi}$ for all $k\in K,
j\not \in K$.
\end{thm}

More can be said about a pair of triples which together span fewer than
five dimensions.

\begin{thm}{Lemma}{twocommon}
Suppose that for some $1\leq l<l'\leq n$ we have 
$A_l \ket{\psi} = A_{l'} \ket{\psi}$ and 
$C_l \ket{\psi} = C_{l'} \ket{\psi}$.
Then $A_k \ket{\psi}$, $B_k \ket{\psi}$, and $C_k \ket{\psi}$
are each orthogonal to
$-i\ket{\psi}$ and to
$A_j \ket{\psi}$, $B_j \ket{\psi},C_j \ket{\psi}$
for all $k\in \{l,l'\},j\not\in\{l,l'\}$.
\end{thm}

\begin{thm}{Proposition}{twocommongen}
{Generalization of~\ref{twocommon}:} Suppose that
$\dim \langle T_l,T_{l'}\rangle \leq 4$ for some $1\leq l<l'\leq n$.
Then $A_k \ket{\psi}$, $B_k \ket{\psi}$, and $C_k \ket{\psi}$ are each
orthogonal to $-i\ket{\psi}$ and to $A_j \ket{\psi}$, $B_j \ket{\psi},C_j
\ket{\psi}$ for all $k\in \{l,l'\},j\not\in\{l,l'\}$.
\end{thm}

We use propositions~\ref{triplesprop},~\ref{mainpropgen}
and~\ref{twocommongen} to establish the lower bound for the rank of
$M$, and therefore also for orbit dimension.

\begin{thm}{Proposition}{minrankM}
{Minimum rank of $M$:} Let $\ket{\psi}\in H$ be a state
vector, and let $M$ be the matrix associated to $\ket{\psi}$.  We
have
$$\rank_{\R} M \geq \left\{
\begin{array}{cc}
\frac{3n}{2} + 1 & \mbox{$n$ even}\\
\frac{3n+1}{2} + 1 & \mbox{$n$ odd}
\end{array}\right..$$
\end{thm}

A product of singlets (together with an unentangled qubit for $n$ odd)
achieves the lower bound established in~\ref{minrankM}.  This
establishes the minimum orbit dimension.

\begin{thm}{Theorem}{minorbthmstmnt}
For the local unitary group
action on state space for $n$ qubits, the smallest orbit dimension is
$$\min\{\dim {\cal O}_x\colon x\in \Proj(H)\} = \left\{
\begin{array}{cc}
\frac{3n}{2} & \mbox{$n$ even}\\
\frac{3n+1}{2} & \mbox{$n$ odd}
\end{array}\right..$$
\end{thm}

This concludes the statements of definitions and results
from~\cite{lyonswalck} to be used in the sequel.

\setcounter{statementnumber}{0}
\section{Further results on ranks of sets of triples}

In this section we develop more facts about ranks of sets of triples of
the columns of the matrix $M$ associated to an $n$-qubit state vector
$\ket{\psi}$ as described in the previous section.  These include
strengthened versions of~\ref{twocommon}, \ref{twocommongen}
and~\ref{minrankM}.  We begin with a general fact about local unitary
invariance.

\begin{thm}
  {Proposition}{ranktripluinv}
The dimension of the span of a union of triples with or without the
rightmost column $-i\ket{\psi}$ is a local unitary
invariant. 
\end{thm}

\begin{proof}
We prove the proposition for the case ``with the rightmost column.''
The same proof works for the case ``without the rightmost column'' by
making the obvious changes.

Let $T_{j_1},T_{j_2},\ldots,T_{j_m}$ be a set of triples of columns of
the matrix $M$ associated to the state vector $\ket{\psi}$.  Let
$U=\prod_{i=1}^n U_i$ be an element of the local unitary group and let
$M'$ with triples
$T_k'=\{A_kU\ket{\psi},B_kU\ket{\psi},C_kU\ket{\psi}\}$ be associated to
the state $\ket{\psi'}=U\ket{\psi}$.  Since $U$ is unitary, the
dimension of the span
$$\dim\langle T_{j_1}'\cup T_{j_2}'\cup \ldots\cup T_{j_m}' \cup
\{-i\ket{\psi'}\}\rangle$$ is equal to the dimension of the span of the set
$$(\ast)\hspace*{.3in}\bigcup_{i=1}^m
\{U^\dag A_{j_i}U\ket{\psi},U^\dag B_{j_i}U\ket{\psi},U^\dag C_{j_i}U\ket{\psi}\}\cup
\{-i\ket{\psi}\}.\hfill \spacer$$
Observe that
\begin{eqnarray*}
U^\dag A_kU &=& (0,\ldots,0,U_k^\dag AU_k,0,\ldots,0)  
=(0,\ldots,0,\mbox{\rm Ad}(U_k^\dag)(A),0,\ldots,0)\\
U^\dag B_kU &=& (0,\ldots,0,U_k^\dag BU_k,0,\ldots,0)
=(0,\ldots,0,\mbox{\rm Ad}(U_k^\dag)(B),0,\ldots,0)\\ 
U^\dag C_kU &=& (0,\ldots,0,U_k^\dag
CU_k,0,\ldots,0)=(0,\ldots,0,\mbox{\rm Ad}(U_k^\dag)(C),0,\ldots,0)   
\end{eqnarray*}
where the zeros occur in all but the $k$th coordinate and $\mbox{\rm
Ad}\colon \mbox{\rm SU}(2)\to \mbox{\rm SO}(\mbox{\rm su}(2))$ denotes the adjoint representation of
$\mbox{\rm SU}(2)$ on its Lie Algebra.  It follows that the span of the
set~$(\ast)$ lies inside the span of the set
$$(\ast\ast)\hspace*{.3in}\bigcup_{i=1}^m
\{ A_{j_i}\ket{\psi},B_{j_i}\ket{\psi}, C_{j_i}\ket{\psi}\}\cup
\{-i\ket{\psi}\}.\hfill \spacer$$
%$$T_{j_1}\cup T_{j_2}\cup \ldots\cup T_{j_m} \cup
%\{-i\ket{\psi}\}$$
and hence that 
$$\dim\langle T_{j_1}'\cup T_{j_2}'\cup \ldots\cup T_{j_m}' \cup
\{-i\ket{\psi'}\}\rangle \leq \dim\langle T_{j_1}\cup T_{j_2}\cup \ldots\cup T_{j_m} \cup
\{-i\ket{\psi}\}\rangle.$$
Reversing the roles of $\ket{\psi}$ and $\ket{\psi'}$ yields that the
span of the set~$(\ast\ast)$ lies inside the span of the set~$(\ast)$,
and hence that 
$$\dim\langle T_{j_1}'\cup T_{j_2}'\cup \ldots\cup T_{j_m}' \cup
\{-i\ket{\psi'}\}\rangle \geq \dim\langle T_{j_1}\cup T_{j_2}\cup \ldots\cup T_{j_m} \cup
\{-i\ket{\psi}\}\rangle.$$
This concludes the proof.
\end{proof}

Next we consider the case of two triples which together span five dimensions.

\begin{thm}{Proposition}{twotripspan5}
Suppose that $\dim \langle T_l,T_{l'}\rangle =5$ for some $1\leq
l<l'\leq n$.  Then there are four independent vectors
$\ket{\zeta_l},\ket{\eta_l} \in T_l$, $\ket{\zeta_{l'}},\ket{\eta_{l'}}
\in T_{l'}$, which are orthogonal to $-i\ket{\psi}$ and to $A_j
\ket{\psi}$, $B_j \ket{\psi},C_j \ket{\psi}$ for all $j\not\in\{l,l'\}$.
\end{thm}

\begin{proof}
Applying the main orthogonality proposition~\ref{mainpropgen}, the
only new claim made in the statement of~\ref{twotripspan5} is that the vectors
$\ket{\zeta_k},\ket{\eta_k} \in T_k$, $k\in \{l,l'\}$ are independent.
In the proof of~\ref{mainpropgen} there is a unitary transformation
$U\colon H\to H$ such that
\begin{eqnarray*}
  U\ket{\zeta_k} &=& B_kU\ket{\psi}\\
  U\ket{\eta_k} &=& C_kU\ket{\psi}
\end{eqnarray*}
for $k\in \{l,l'\}$ and $A_lU\ket{\psi}$ is collinear with
$A_{l'}U\ket{\psi}$.  Since $\dim\langle T_l,T_{l'}\rangle = 5$, the
span of $\displaystyle \{A_kU \ket{\psi}$, $B_k U\ket{\psi},C_k
U\ket{\psi}\colon k=l,l'\}$ is also five dimensional
by~\ref{ranktripluinv}.  Thus the collinearity of $A_lU\ket{\psi}$ and
$A_{l'}U\ket{\psi}$ implies that $U\ket{\zeta_l}$, $U\ket{\eta_l}$,
$U\ket{\zeta_{l'}}$, and $U\ket{\eta_{l'}}$ are independent, and
therefore that $\ket{\zeta_l}$, $\ket{\eta_l}$, $\ket{\zeta_{l'}}$, and
$\ket{\eta_{l'}}$ are independent.
\end{proof}

Next, a small observation proves a stronger version of~\ref{twocommon}.

\begin{thm}{Lemma}{twocommonstrong}
Suppose that for some $1\leq l<l'\leq n$ we have $A_l \ket{\psi} =
A_{l'} \ket{\psi}$ and $C_l \ket{\psi} = C_{l'} \ket{\psi}$.  Then $B_l
\ket{\psi} = -B_{l'} \ket{\psi}$, the dimension of $\langle
T_l,T_{l'}\rangle$ is three, and $\langle T_l,T_{l'}\rangle$ is
orthogonal to $-i\ket{\psi}$ and to $A_j \ket{\psi}$, $B_j
\ket{\psi},C_j \ket{\psi}$ for all $j\not\in\{l,l'\}$.
\end{thm}

\begin{proof}
We only need to prove that $B_l\ket{\psi} = -B_{l'} \ket{\psi}$.  The
rest of the statement follows from~\ref{twocommon}.
Equation~(\ref{akact}) and the hypothesis $A_l \ket{\psi} = A_{l'}
\ket{\psi}$ imply that $(-1)^{i_l}c_I = (-1)^{i_{l'}}c_I$, so if
$c_I\neq 0$ then $i_l = i_{l'}$ mod~2.  It follows that if $c_{I_l}\neq
0$ then $i_l = i_{l'}+1$ mod~2.  Equation~(\ref{ckact}) and the
hypothesis $C_l \ket{\psi} = C_{l'} \ket{\psi}$ imply that
$c_{I_l}=c_{I_{l'}}$ for all $I$.  Hence, for all $I$ we have
$$\bra{I} B_l \ket{\psi} = (-1)^{i_l}c_{I_l} = (-1)^{i_{l'}+1}c_{I_{l'}}
= -\bra{I} B_{l'}\ket{\psi}.$$
This establishes the claim.
\end{proof}

The strengthened Lemma~\ref{twocommonstrong} yields the
following strengthened version of~\ref{twocommongen}. We omit the
proof because it requires only a minor change to
apply~\ref{twocommonstrong} in the proof of~\ref{twocommongen}.

\begin{thm}{Proposition}{twocommonstronggen}
%{Two triples which span 3 or 4 dimensions, stronger version} 
{Strengthened version of~\ref{twocommongen}:} Let $\ket{\psi}$ be a
state vector and let $M$ be its associated matrix.  Suppose that $\dim
\langle T_l,T_{l'}\rangle \leq 4$ for some $1\leq l<l'\leq n$.  Then
$\ket{\psi}$ is local unitary equivalent to a state vector $\ket{\psi'}$
such that $A_l \ket{\psi'} =A_{l'} \ket{\psi'}$, $B_l \ket{\psi'}
=-B_{l'} \ket{\psi'}$, $C_l \ket{\psi'} =C_{l'} \ket{\psi'}$, the
dimension of $\langle T_l,T_{l'}\rangle$ is three, and $\langle
T_l,T_{l'}\rangle$ is orthogonal to $-i\ket{\psi}$ and to $A_j
\ket{\psi}$, $B_j \ket{\psi},C_j\ket{\psi}$ for all $j\not\in\{l,l'\}$.
\end{thm}

Next we state and prove a stronger version of~\ref{minrankM}.

\begin{thm}{Proposition}{minrankMstrong}
{Minimum rank for submatrices of $M$:}
Let ${\cal S}=T_{i_1}\cup T_{i_2}\cup \cdots \cup T_{i_q}\cup
\{-i\ket{\psi}\}$ be a union of $q$ triples together with the rightmost
column $-i\ket{\psi}$ of $M$.  Then 
$$\displaystyle \dim \langle {\cal S}\rangle
\geq  \left\{ \begin{array}{cc} \frac{3q}{2}+1 & \mbox{$q$ even}\\
\frac{3q+1}{2}+1 & \mbox{$q$ odd}\end{array}\right. .$$
\end{thm}

\begin{proof}
Let ${\cal S}_0\subseteq {\cal S}$ be a union of some
number $p$ of triples, maximal with respect 
to the property that $\langle {\cal S}_0 \rangle$ contains a subspace
$W$ for which
\begin{enumerate}
\item[(i)] \label{propa} $\displaystyle \dim W
\geq  \left\{ \begin{array}{cc} \frac{3p}{2} & \mbox{$p$ even}\\
\frac{3p+1}{2} & \mbox{$p$ odd}\end{array}\right.$,\hspace*{.25in}
and
\item[(ii)] \label{propb} $W \perp \langle{\cal
C}\setminus{\cal S}_0 \rangle$.
\end{enumerate}

We separate the argument into cases.  We show that in every case,
either~\ref{minrankMstrong} holds or we can derive a contradiction by
constructing a superset ${\cal S}_1$ such that ${\cal S}_0
\subseteq{\cal S}_1 \subseteq {\cal S}$, and ${\cal S}_1$ is the union
of some number $p'>p$ triples and contains a subspace $W'$ satisfying
properties~(i) and ~(ii) with $p'$ in place of $p$.  The construction of
${\cal S}_1$ violates the maximality of ${\cal S}_0$ and therefore rules
out the case in question.

Case 1: Suppose that $p=q$.  Then~\ref{minrankMstrong} holds.

Case 2: Suppose that $p<q$ and that the triples
$T_{j_1},T_{j_2},\ldots,T_{j_{q-p}}$ in ${\cal S}\setminus {\cal S}_0$
have the maximum possible span, that is, 
$$\dim \langle T_{j_1},T_{j_2},\ldots,T_{j_{q-p}}\rangle =3(q-p).$$
Properties~(i) and~(ii) imply that 
\begin{eqnarray*}
\dim\langle {\cal S}\rangle &\geq& \dim W + \dim \langle{\cal S}\setminus{\cal S}_0 \rangle
\\
&\geq& \frac{3p}{2} + 3(q-p)\\
&=& \frac{6q-3p}{2}\\
&\geq& \frac{6q - (3q-3)}{2}\hspace*{.2in}\mbox{(since $p\leq q-1$)}\\
&=& \frac{3q+3}{2}\\
&=& \frac {3q+1}{2} + 1
\end{eqnarray*}
and so~\ref{minrankMstrong} holds.
Note that if $p=q-1$, the hypothesis of full span is met
by~\ref{triplesprop}.  Therefore in the remaining cases we need only
consider $p\leq q-2$.

Case 3: Suppose $p\leq q-2$ and that there is a pair of triples
$T_l,T_{l'}$ in ${\cal S}\setminus{\cal S}_0$ with
$\dim\langle T_l,T_{l'}\rangle \leq 4$.  Let ${\cal S}_1 = {\cal S}_0
\cup T_l\cup T_{l'}$, let $p'=p+2$, and let $W'=W \oplus \langle T_l\cup
T_{l'} \rangle$, where ``$\oplus$'' denotes the orthogonal direct sum.
That the sum is orthogonal is guaranteed by property~(ii) for $W$.
Proposition~\ref{twocommongen} implies that property~(ii) also holds
for the pair $({\cal S}_1,W')$ and that $\dim W' \geq \dim W + 3$ (in
fact, we have equality by~\ref{twocommonstronggen}).  It follows that
if $p$ is even, so is $p'$ and we have
$$\dim W' \geq \frac{3p}{2} +3 = \frac{3p+6}{2} = \frac{3(p+2)}{2} =
\frac{3p'}{2}$$
and similarly if $p$ and $p'$ are odd we have
$$\dim W' \geq \frac{3p+1}{2} +3 = \frac{3p'+1}{2}$$
so $({\cal S}_1, W')$ satisfies property~(i).  Thus ${\cal S}_1$ violates the
maximality of ${\cal S}_0$, so we conclude that the hypothesis of case~3
is impossible.

Case 4: Suppose $p\leq q-2$ and that there is a pair of triples
$T_l,T_{l'}$ in ${\cal S}\setminus{\cal S}_0$ 
such that $\dim\langle T_l,T_{l'}\rangle =5$.
Applying~\ref{twotripspan5} we have four independent vectors
$$\ket{\zeta_l},\ket{\eta_l}\in \langle T_l\rangle, \hspace*{.25in}
\ket{\zeta_{l'}},\ket{\eta_{l'}}\in \langle T_{l'} \rangle$$ orthogonal
to all column vectors of $M$ in columns outside of triples $T_l,T_{l'}$,
so once again ${\cal S}_1 = {\cal S}_0 \cup T_l\cup T_{l'}$ with the
subspace
$$W'=W\oplus \langle \ket{\zeta_l},\ket{\eta_l}, \ket{\zeta_{l'}},\ket{\eta_{l'}}\rangle $$
violates the maximality of ${\cal S}_0$.  We conclude that the
hypothesis of case~4 is impossible.  

Case 5: The only remaining possibility is that $p\leq q-3$.
Let ${\cal T}=\{T_{j_1},T_{j_2},\ldots,T_{j_m}\}$ be a set of
triples in ${\cal S}\setminus {\cal S}_0$ with $m\geq 3$ minimal with
respect to the property 
$$\dim \langle T_{j_1},T_{j_2},\ldots,T_{j_m}\rangle < 3m.$$
Applying~\ref{mainpropgen} we have 2 vectors
$$\ket{\zeta_{k}},\ket{\eta_{k}}\in \langle T_{k}\rangle$$
for each of the $m'\geq 2$ elements $k\in K$.  Let 
$${\cal S}_1 = {\cal S}_0 \cup \left(\bigcup_{k\in K} T_k\right),$$
let $p'=p+m'$, and let 
$$W'=W \oplus \langle \{\ket{\zeta_{k}},\ket{\eta_{k}}\}_{k\in
  K}\rangle.$$ Note that property~(ii) holds for $({\cal S}_1,W')$.  If
  $m'<m$, then the $2m'$ vectors in
  $\{\ket{\zeta_{k}},\ket{\eta_{k}}\}_{k\in K} $ are independent by the
  minimality of ${\cal T}$, so we have
$$\dim W' \geq \dim W + 2m' \geq \frac{3p}{2} + 2m' =
  \frac{3p'+m'}{2}\geq \frac{3p'+1}{2} $$
so property~(i) holds for $({\cal S}_1,W')$, but this contradicts the
  maximality of ${\cal S}_0$.  Finally, if $m'=m$, then $m\geq 4$ (since
  $m'$ is even) and at least $2(m-1)$ of
  the vectors in $\{\ket{\zeta_{k}},\ket{\eta_{k}}\}_{k\in K} $ must be
  independent, again by the minimality of ${\cal T}$.  If $p$ is even,
  then $p'=p+m$ is also even and we have
$$\dim W' \geq \dim W + 2(m-1) \geq \frac{3p}{2} + 2(m-1) =
  \frac{3p'+m-4}{2}\geq \frac{3p'}{2}.$$  
If $p$ is odd, then $p'=p+m$ is odd and we have
$$\dim W' \geq \dim W + 2(m-1) \geq \frac{3p+1}{2} + 2(m-1) =
  \frac{3p'+m-3}{2}\geq \frac{3p'+1}{2}.$$  
Thus ${\cal S}_1$ with the subspace $W'$ violates the maximality
  of ${\cal S}_0$.  We conclude that the hypothesis of case~5 is impossible.

Having exhausted all possible cases, this completes the proof
of~\ref{minrankMstrong}.
\end{proof}

Next we state and prove a general statement about additivity of ranks for
bipartite states.

\begin{thm}
  {Proposition}{bipartiteranksadd} Let $\ket{\psi}=\ket{\psi_1}\otimes
\ket{\psi_2}$ be a state vector for a bipartite state, where
$\ket{\psi_j}$ is an $n_j$-qubit state vector for $j=1,2$.  Let $M$ be
the matrix associated to $\ket{\psi}$, and let $M_j$ denote the
associated matrix for $\ket{\psi_j}$ for $j=1,2$.
Let ${\cal S}_1$ and ${\cal S}_2$ be the following submatrices of $M$
\begin{eqnarray*}
  {\cal S}_1 &=& T_1\cup T_2 \cup \cdots \cup T_{n_1}\cup
  \{-i\ket{\psi}\}\\
  {\cal S}_2 &=& T_{n_1+1}\cup T_{n_2+2} \cup \cdots \cup T_{n_1+n_2}\cup \{-i\ket{\psi}\}
\end{eqnarray*}
and let ${\cal B'} = T'_{j_1}\cup T'_{j_2}\cup \ldots \cup T'_{j_m}$ be a
union of triples in $M_1$ with corresponding union ${\cal B} =
T_{j_1}\cup T_{j_2}\cup \ldots \cup T_{j_m}$ in $M$.
We have
\begin{enumerate}
%  \item[(i)] $Gx\approx  G_1x_1\times G_2x_2$, where ``$\approx$'' denotes homeomorphism,
\item[(i)]  $\rank_\R M - 1 = (\rank_\R M_1 - 1) +
(\rank_\R M_2 - 1)$, 
\item[(ii)] $\rank_\R M_j = \dim\langle{\cal S}_j\rangle$ for $j=1,2$, and
\item[(iii)] $\dim\langle {\cal B'}\rangle = \dim\langle {\cal B}\rangle.$
\end{enumerate}
\end{thm}

\begin{proof}
Let $x$ denote the state represented by $\ket{\psi}$ and let $G$ denote
the local unitary group.  Let $x_j$
denote the state represented by $\ket{\psi_j}$, and let $G_j$ denote the local
unitary group for $j=1,2$, so we have $G=G_1\times G_2$.

It is easy to see that the $G$-orbit of $x$ is diffeomorphic to the product
of the $G_j$-orbits of the $x_j$, so dimensions add.
$$\dim Gx =  \dim G_1x_1 +  \dim G_2x_2$$
Applying~\ref{orbitdimasrank}, it follows immediately that~(i) holds.

For~(ii), observe that the columns of ${\cal S}_j$ are simply the
comlumns of $M_j$ tensored with $\ket{\psi_2}$.  The same reasoning
applied to the subset ${\cal B}\subseteq {\cal S}_1$ establishes~(iii).
\end{proof}

We end this section with statements about factoring singlets and
unentangled qubits.

\begin{thm}{Proposition}{twotripspan3factors}
%{Factoring singlets:} 
There are two triples $T_l,T_{l'}$ with $\dim \langle T_l,T_{l'}\rangle
= 3$ if and only if the state factors as a product of a singlet in
qubits $l,l'$ times a state in the remaining qubits.
\end{thm}

\begin{proof}
Without loss of generality, let us renumber the qubits so that $l=1,l'=2$.  

First we prove the ``if'' part of the statement.  Let $\ket{\psi}=\ket{s}\otimes
\ket{\phi}$ where $\ket{s}$ is a singlet in qubits 1 and 2, and
$\ket{\phi}$ is an $(n-2)$-qubit state.  Then $\ket{\psi}$ is local
unitary equivalent to $\ket{\psi'}= \ket{s'}\otimes \ket{\phi}$, where
$\ket{s'}=\ket{00}+\ket{11}$.   A simple calculation shows that
the two triples in the matrix for a 2-qubit singlet state vector
$\ket{s'}$ together span three dimensions.  Therefore
by~\ref{bipartiteranksadd}~(iii), the dimension of the span of triples
1 and 2 in the matrix for $\ket{\psi'}$ is also three.  Since the
rank of unions of triples is local unitary invariant
by~\ref{ranktripluinv}, we conclude that the dimension of the span of
triples 1 and 2 in the matrix for $\ket{\psi}$ is also three.

Next we prove ``only if.''  Let $\ket{\psi}$ be a state vector for which
$\dim \langle T_1,T_{2}\rangle = 3$.  By~\ref{twocommonstronggen},
$\ket{\psi}$ is local unitary equivalent to state vector $\ket{\psi'}$
for which $A_1\ket{\psi'}=A_{2}\ket{\psi'}$ and
$C_1\ket{\psi'}=C_{2}\ket{\psi'}$.  Equation~(\ref{akact}) and the
hypothesis $A_1 \ket{\psi} = A_{2} \ket{\psi}$ imply that $(-1)^{i_1}c_I
= (-1)^{i_{2}}c_I$, so if $c_I\neq 0$ then $i_1 = i_{2}$ mod~2, so every
$I$ for which $c_I\neq 0$ has either both zeros or both ones in the
first two indices. Equation~(\ref{ckact}) and the hypothesis $C_1
\ket{\psi} = C_{2} \ket{\psi}$ imply that $c_{I_1}=c_{I_{2}}$ for all
$I$.  Apply this to $I$ for which $i_1\neq i_2$ and we get
$$c_{(00i_3i_4\ldots i_n)} = c_{(11i_3i_4\ldots i_n)}$$
for all $(i_3i_4\ldots i_n)$.  It follows that $\ket{\psi'}$ factors as
a product
$$\ket{\psi'} = (\ket{00}+\ket{11})\otimes \ket{\phi}$$
where $\ket{\phi}$ is an $(n-2)$-qubit state.  Therefore $\ket{\psi}$ is
a product of a singlet state in the first two qubits times a state in
the remaining qubits.
\end{proof}

\begin{thm}
  {Lemma}{trippluslonelyspan3lemma}
If $A_j\ket{\psi}=i\ket{\psi}$ then the $j$th qubit is unentangled.
\end{thm}

\begin{proof}
Renumber the qubits, if necessary, so that $j=1$.  By~(\ref{akact}), the
hypothesis $A_1\ket{\psi}=i\ket{\psi}$ implies that if $c_I\neq 0$ then
$i_1=0$.  Therefore $\ket{\psi}$ factors as a product
$$\ket{\psi}= \ket{0}\otimes \ket{\phi}$$ where $\ket{\phi}$ is an
$(n-1)$-qubit state vector.
\end{proof}

\begin{thm}
  {Proposition}{trippluslonelyspan3} If the dimension of the span of a
triple together with the rightmost column $-i\ket{\psi}$ is three, then
the state factors as an unentangled qubit times a state in the remaining
qubits.
\end{thm}

\begin{proof}
Let $T_j$ be a triple such that $\dim \langle T_j \cup
\{-i\ket{\psi}\}\rangle=3$.  Since $i\ket{\psi}$
lies in the span of $\langle{T_j}\rangle$, we may write
$$i\ket{\psi} = \alpha A_{j}\ket{\psi} + \beta B_{j}\ket{\psi} + \gamma
C_{j}\ket{\psi}$$ for some real $\alpha,\beta,\gamma$.  Choose $R\in
\mbox{\rm SO}(\mbox{\rm su}(2))$ such that
$$R(A) = \alpha A + \beta B + \gamma C.$$ Since the adjoint
representation $\mbox{\rm Ad}\colon \mbox{\rm SU}(2)\to \mbox{\rm
SO}(\mbox{\rm su}(2))$ is surjective, we can choose $U_{j}\in \mbox{\rm
SU}(2)$ such that $\mbox{\rm Ad}(U_{j}^\dag) = R$, that is, $U_{j}^\dag
X U_{j} = R(X)$ for all $X\in \mbox{\rm su}(2)$.  For $1\leq k \leq n$,
$k\neq j$, set $U_k$ equal to the identity.  Finally, let $U\in
G=\mbox{\rm SU}(2)^n$ be $U=\prod_{i=1}^n U_i$.  We have
$$U^\dag A_j U\ket{\psi} = i\ket{\psi}.$$
Applying $U$ to both sides, we obtain
$$A_jU\ket{\psi} = iU\ket{\psi}.$$ Applying
Lemma~\ref{trippluslonelyspan3lemma} to the matrix $M'$ for the state
$U\ket{\psi}$ shows that the $j$th qubit is unentangled for the state
$U\ket{\psi}$.  Since unentanglement of a particular qubit is local
unitary invariant, the proposition is established.
\end{proof}

\begin{thm}{Proposition}{unentrank}
{Rank of unentangled triples:}
If a state has $k$ unentangled qubits, the rank of the union of the
triples corresponding to those qubits together with the rightmost column
$-i\ket{\psi}$ of $M$ is $2k+1$. 
\end{thm}

\begin{proof}
Let $\ket{\psi}$ be a state vector for a state with $k$ unentangled
qubits, and let us renumber the qubits, if necessary, so that the
unentangled qubits are numbered 1 through $k$.  The state vector
$\ket{\psi}$ is local unitary equivalent to a state vector
$$\ket{\psi'} = \ket{00\cdots 0}\otimes \ket{\phi}$$ where
$\ket{00\cdots 0}$ is the product of $k$ unentangled qubits and
$\ket{\phi}$ is an $(n-k)$-qubit state.  Let $M'$ be the matrix
associated to $\ket{\psi'}$, $M_1$ the matrix for $\ket{00\cdots 0}$,
and $M''$ the matrix for the single qubit state $\ket{0}$.
Apply~\ref{bipartiteranksadd}~(i) to $\ket{00\cdots 0}=\ket{0}\otimes
\cdots \otimes \ket{0}$ to get $\rank_\R M_1 = 2k+1$ (using the fact
that the single qubit state $\ket{0}$ has $\rank_\R M'' = 3$).  Then
apply~\ref{bipartiteranksadd}~(ii) to the set ${\cal S}_1$ which is
the union of the first $k$ triples of the matrix $M'$ together with the
column $-i\ket{\psi'}$ to get $\dim\langle {\cal S}_1\rangle = \rank_\R
M_1 = 2k+1$.  Finally, apply~\ref{ranktripluinv} to conclude that the
desired statement holds for $\ket{\psi}$.
\end{proof}

\setcounter{statementnumber}{0}
\section{Minimum dimension orbit classification}

Now we prove that any state with minimum orbit dimension is a product of
singlets for $n$ even, times an unentangled qubit for $n$ odd.

\begin{thm}
  {Main Lemma}{minrankimpliessinglet}
For $n\geq 2$, if $M$ has minimum rank, then there is some pair of triples whose span
is three dimensional.
\end{thm}

\begin{proof}
Suppose not.  A consequence of~\ref{twocommonstronggen} is that every
pair of triples spans either three, five or six dimensions.  We can rule out the
possibility that some pair of triples spans five dimensions, as follows.
If there is a pair $T_l,T_{l'}$ of triples which spans five dimensions,
then by~\ref{twotripspan5}, the pair contributes four independent column
vectors which are orthogonal to every column vector in the set ${\cal
S}={\cal C}\setminus (T_l\cup T_{l'})$.  Applying~\ref{minrankMstrong}
to ${\cal S}$ with with $q=n-2$, we have
\begin{eqnarray*}
\rank M &\geq&  4 + \left\{ \begin{array}{cc} \frac{3q}{2}+1 & \mbox{$n$ even}\\
\frac{3q+1}{2}+1 & \mbox{$n$ odd}\end{array}\right.\\
 &=& \left\{ \begin{array}{cc} \frac{3n+2}{2}+1&\mbox{$n$ even}\\
\frac{3n+3}{2}+1& \mbox{$n$ odd}\end{array}\right.   
\end{eqnarray*}
which is greater than minimum, and therefore impossible.

Thus we need only consider the case where every pair of triples spans six
dimensions.  

Let ${\cal T}=\{T_{j_1},T_{j_2},\ldots,T_{j_m}\}$ be a set of $m$
triples minimal with respect to the property
$$\dim \langle T_{j_1},T_{j_2},\ldots,T_{j_m}\rangle < 3m$$
(``minimal'' means that ${\cal T}$ contains no proper subset of triples
which satisfy the given property; thus, any subset of $m'<m$ triples of
${\cal T}$ has ``full'' span of $3m'$ dimensions).
We know such a ${\cal T}$ exists since the rank of $M$ is minimum.
Apply~\ref{mainpropgen} to get a subset
$K\subseteq\{j_1,j_2,\ldots,j_m\}$ with some
even positive number $m'$ of elements and vectors
$\ket{\zeta_k},\ket{\eta_k}$ in $T_k$ for $k\in K$.  Let $\displaystyle
{\cal T}'=\bigcup_{k\in K} T_k$ and let ${\cal S}={\cal C}\setminus {\cal T}'$ be
the union of the $q=n-m'$ triples not in ${\cal T}'$ together with the
rightmost column $-i\ket{\psi}$ of $M$.  

If $m'<m$, the minimality of ${\cal T}$ guarantees that the $2m'$ vectors
$\{\ket{\zeta_{k}},\ket{\eta_{k}}\}_{k\in K} $ are independent, so the
rank of $M$ is at least (apply~\ref{minrankMstrong} to ${\cal S}$ with
$q=n-m'$)
\begin{eqnarray*}
\rank M &\geq&  2m' + \left\{ \begin{array}{cc} \frac{3q}{2}+1 & \mbox{$n$ even}\\
\frac{3q+1}{2}+1 & \mbox{$n$ odd}\end{array}\right.\\
 &=& \left\{ \begin{array}{cc} \frac{3n+m'}{2}+1&\mbox{$n$ even}\\
\frac{3n+1+m'}{2}+1& \mbox{$n$ odd}\end{array}\right.   
\end{eqnarray*}
which is greater than minimum, so this case cannot occur.

If $m'=m$, the minimality of ${\cal T}$ guarantees that at least $2(m'-1)$ of
the vectors in $\{\ket{\zeta_{k}},\ket{\eta_{k}}\}_{k\in K} $ are
independent, so the rank of $M$ is at least 
\begin{eqnarray*}
\rank M &\geq&  2(m'-1) + \left\{ \begin{array}{cc} \frac{3q}{2}+1 & \mbox{$n$ even}\\
\frac{3q+1}{2}+1 & \mbox{$n$ odd}\end{array}\right.\\
 &=& \left\{ \begin{array}{cc} \frac{3n+m'-4}{2}+1&\mbox{$n$ even}\\
\frac{3n+1+m'-4}{2}+1& \mbox{$n$ odd}\end{array}\right..   
\end{eqnarray*}
If $m'>4$, this is greater than minimum, so we may assume that $m'=m\leq
4$.  

We rule out the possibility $m=2$ since every pair of triples spans six
dimensions, so the only remaining case is $m'=m=4$.  We may further
assume that {\em every} minimal set of triples that has less than full span
consists of $m=4$ triples, and that applying~\ref{mainpropgen} to such
a set yields $m'=4$.  Thus the columns of $M$ decompose into a disjoint
union
$${\cal C} = {\cal S}_0 \cup {\cal T}_1 \cup {\cal T}_2 \cup \cdots \cup
{\cal T}_t$$ where each ${\cal T}_i$ is a union of four triples for which
applying~\ref{mainpropgen} yields six independent vectors orthogonal to
$\langle {\cal C}\setminus {\cal T}_i\rangle$, and ${\cal S}_0$ is a union of
$q=n-4t$ triples together with the rightmost column $-i\ket{\psi}$ of
$M$ such that the span of the union of the triples in ${\cal S}_0$ is
$3q$ dimensional.

We consider the cases $q \geq 2$, $q=0$, and finally $q=1$.

If $q\geq 2$ then we have 
$$\rank M \geq 6t + 3q =
6\left(\frac{n-q}{4}\right)+3q=\frac{3n+3q}{2}$$
which is greater than minimum, so this case cannot occur.

If $q=0$ or $q=1$, let ${\cal S}_1={\cal S}_0 \cup {\cal T}_1$ and consider the
disjoint union
$${\cal C} = {\cal S}_1 \cup {\cal T}_2 \cup \cdots \cup
{\cal T}_t$$
of the set ${\cal S}_1$ with  $t'=t-1$ unions of four triples $\{{\cal
  T}_i\}_{i=2}^t$.

If $q=0$, ${\cal S}_1$ has a span of at least nine dimensions since any
three of the triples in ${\cal T}_1$ have full span.  We have
$$\rank M \geq 9 + 6t' = 9 + 6\left(\frac{n-4}{4}\right) =
\frac{3n+6}{2}$$
which is greater than minimum, so this case cannot occur.

If $q=1$, then ${\cal S}_1$ is the union of five triples together with
the rightmost column $-i\ket{\psi}$.
%, say $T_{j_1},T_{j_2},\ldots,T_{j_5}$.   
If any subset
of four of those five triples has full span, then we have
$$\rank M \geq 12 + 6t' = 12 + 6\left(\frac{n-5}{4}\right) =
\frac{3n+9}{2}$$
which is greater than minimum, so it must be the case that {\em all}
subsets of four triples have less than full span, so any subset of four
triples contributes six independent vectors orthogonal to the remaining
triple and the rightmost column $-i\ket{\psi}$ of $M$.  If any one of the
qubits corresponding to one of the five triples is {\em not} unentangled, then
by~\ref{trippluslonelyspan3} we have
$$\rank M \geq 6 + 4 + 6t' = 10 + 6\left(\frac{n-5}{4}\right) =
\frac{3n+5}{2}$$
which is greater than minimum, so it must be the case that {\em all} five
qubits are unentangled.  But then by~\ref{unentrank} we have
$$\rank M \geq 11 + 6t' = 11 + 6\left(\frac{n-5}{4}\right) =
\frac{3n+7}{2}$$
which is greater than minimum.
 
Since all possible cases lead to contradictions, we conclude that some
pair of triples must have a three dimensional span.
\end{proof}

\begin{thm}
 {Corollary}{decompcor} Any state which has minimum orbit dimension is a
product of singlets when $n$ is even, together with an unentangled
qubit when $n$ is odd.
\end{thm}

\begin{proof}
Let $\ket{\psi}$ be a state vector for a state
with minimum orbit dimension, with associated matrix $M$.  Apply
Lemma~\ref{minrankimpliessinglet} to $M$ to get a pair of triples whose
span is three dimensional.  By~\ref{twotripspan3factors}, $\ket{\psi}$
factors as a product of a singlet in those two qubits times an
$(n-2)$-qubit state, say, with state vector $\ket{\psi_1}$, in the
remaining qubits.  Let $M_1$ be the matrix associated to
$\ket{\psi_1}$.  By~\ref{bipartiteranksadd}~(i), $M_1$ also has
minimum rank.  Repeating this reasoning yields a sequence
$\ket{\psi},\ket{\psi_1},\ket{\psi_2},\ldots$ which eventually exhausts
all the qubits of $\ket{\psi}$ unless $n$ is odd, in which case a single
unentangled qubit remains.
\end{proof}

Next we classify states with minimum orbit dimension up to local unitary
equivalence. 

\begin{thm}{Proposition}{sepsingprod}
{Separation of singlet products:}
Products of singlets are local unitary equivalent if and only if the choices of
entangled pairs are the same in each product.
\end{thm}

\begin{proof}
If $\ket{\psi}$ is a product of singlets for which some
pair of qubits, say qubits 1 and 2, forms a singlet, then $\ket{\psi}$
is of the form $\ket{\alpha}\otimes \ket{\beta}$,
where $\ket{\alpha}$ is a singlet in qubits 1 and 2.  Clearly, any state
local unitary equivalent to $\ket{\psi}$ is also of this form.
\end{proof}

The following classification theorem summarizes the results of this section.

\begin{thm}{Theorem}{minorbclassthm}
{Classification of states with minimum orbit dimension:} An
$n$-qubit pure state has minimum orbit dimension $3n/2$ ($n$ even) or
$(3n+1)/2$ ($n$ odd) if and only if it is a product of singlets, together
with an unentangled qubit for $n$ odd.  Furthermore, two of these states
which do not have the same choices for pairs of entangled qubits are not
local unitary equivalent.
\end{thm}

\section{Conclusion}

The classification of types of quantum entanglement is a difficult but
central problem in the field of quantum information.
Entanglement types partially distinguish themselves by
their local unitary orbit and their orbit dimension.  As an integer
that can be readily calculated for a given quantum state, orbit
dimension is a convenient LU invariant.  It provides a useful ``first
stratification'' of quantum state space, which suggests a two-step
program for entanglement classification.  The first step is to
understand the possible orbit dimensions for a composite quantum system, and
the second step is to understand the types of entanglement that occur
in each orbit dimension.  For pure states of $n$-qubits, the first
step was achieved in~\cite{lyonswalck}.  The present work completes the second
step for the orbits of minimum dimension.  In particular, states with
minimum orbit dimension are precisely products of pairs of qubits,
each pair in a singlet state (or LU equivalent).  While there is much
work left to be done in the second step of the classification program,
it is worth remarking that the present results, dealing with an
arbitrary number of qubits, give some hope that a meaningful
classification of entanglement for $n$ qubits is possible.

Reduced orbit dimension states appear to be the most interesting states.
Carteret and Sudbery~\cite{carteret00a}
pointed out that reduced orbit dimension
states (which they call ``exceptional states'') must have extreme values
of the local unitary invariants that are used in the construction
of all measures of entanglement.  They concluded that reduced orbit
dimension states should be expected to be particularly interesting
and important.
Subsequent studies have confirmed this expectation.
All of the ``famous'' states that theorists use for examples
and that experimentalists try to exhibit in the laboratory
have reduced orbit dimension (examples include
the EPR singlet state, the GHZ state, unentangled states, the W state,
and $n$-cat states).
Both the most entangled and the least entangled states states have
reduced orbit dimension.

Orbit dimension provides a useful first step in entanglement
classification, but the numerical value of orbit dimension,
beyond providing a sense of how rare an entanglement type is,
does not carry a simple physical meaning.
For example, in the case of pure three-qubit
states~\cite{carteret00a,walck},
the minimum orbit dimension is 5 and the
maximum orbit dimension is 9.  States with orbit dimension 5
consist only of products of a singlet pair and a qubit,
orbit dimension 6 contains the unentangled states,
orbit dimension 7 contains GHZ states
as well as products of generic two-qubit states
with an unentangled qubit, orbit dimension 8 contains the W state,
and orbit dimension 9 contains all generic states.

Minimum orbit dimension states have the most symmetry with respect
to local transformations in that they remain invariant under a larger
class of transformations than any other states.  They are
maximal symmetry generalizations of the spin singlet state, which is invariant
(as a quantum state, not an entanglement type)
to any rotation applied identically to both spins.
The present result, that the $n$-qubit maximal symmetry generalizations
of the two-qubit singlet state are themselves products of singlets,
shows a special role for the two-qubit singlet state in the theory
of $n$-qubit quantum entanglement.

Linden, Popescu, and Wootters~\cite{linden02a,linden02b}
have shown that almost all pure $n$-qubit quantum
states lack essential $n$-qubit quantum entanglement
in the sense that they can be reconstructed from their reduced density
matrices.
It appears likely
(and is known in the three-qubit case) that states with essential
$n$-qubit entanglement are found among the reduced orbit dimension
states.
The present work shows that minimum orbit dimension states
do not exhibit essential $n$-qubit entanglement
for $n \geq 3$.  Rather, minimum
orbit dimension states maximize pairwise entanglement.
We conjecture that states with
minimum orbit dimension among \emph{non-product} states
have essential $n$-qubit entanglement in the sense
of~\cite{linden02a,linden02b}.

\section*{Acknowledgments}
We thank the anonymous referees for their helpful suggestions.
Co-author Walck thanks the Research Corporation for their support.

\bibliographystyle{unsrt}

%\bibliography{arxbib}

\begin{thebibliography}{10}

\bibitem{nielsenchuang}
Michael~A. Nielsen and Isaac~L. Chuang.
\newblock {\em Quantum Computation and Quantum Information}.
\newblock Cambridge University Press, 2000.

\bibitem{gudder03}
Stan Gudder.
\newblock Quantum computation.
\newblock {\em Amer. Math. Monthly}, 110(3):181--201, 2003.

\bibitem{linden98}
N.~Linden and S.~Popescu.
\newblock On multi-particle entanglement.
\newblock {\em Fortschr. Phys.}, 46:567--578, 1998.
\newblock e-print quant-ph/9711016.

\bibitem{linden99}
N.~Linden, S.~Popescu, and A.~Sudbery.
\newblock Non-local properties of multi-particle density matrices.
\newblock {\em Phys. Rev. Lett.}, 83:243--247, 1999.
\newblock e-print quant-ph/9801076.

\bibitem{acin00}
A.~Ac{\'{\i}}n, A.~Andrianov, L.~Costa, E.~Jan{\'{e}}, J.~I. Latorre, and
  R.~Tarrach.
\newblock Generalized {S}chmidt decomposition and classification of
  three-quantum-bit states.
\newblock {\em Phys. Rev. Lett.}, 85:1560, 2000.
\newblock e-print quant-ph/0003050.

\bibitem{acin01}
A.~Ac{\'{\i}}n, A.~Andrianov, E.~Jan{\'{e}}, and R.~Tarrach.
\newblock Three-qubit pure-state canonical forms.
\newblock {\em J. Phys. A}, 34:6725, 2001.
\newblock e-print quant-ph/0009107.

\bibitem{carteret00a}
H.~A. Carteret and A.~Sudbery.
\newblock Local symmetry properties of pure 3-qubit states.
\newblock {\em J. Phys. A}, 33:4981--5002, 2000.
\newblock e-print quant-ph/0001091.

\bibitem{lyonswalck}
David~W. Lyons and Scott~N. Walck.
\newblock Minimum orbit dimension for local unitary action on $n$-qubit
pure states.
\newblock {\em J. Math. Phys.} 46(10):102106, 2005.
\newblock e-print quant-ph/0503052

\bibitem{walck}
Scott~N. Walck, James~K. Glasbrenner, Matthew~H. Lochman, and Shawn~A. Hilbert.
\newblock Topology of the three-qubit space of entanglement types.
\newblock {\em Phys. Rev. A} 72:052324, 2005.
\newblock e-print quant-ph/0507208

\bibitem{linden02a}
N. Linden, S. Popescu, and W.~K. Wootters.
\newblock Almost Every Pure State of Three Qubits Is Completely
Determined by Its Two-Particle Reduced Density Matrices.
\newblock {\em Phys. Rev. Lett.} 89:207901, 2002.

\bibitem{linden02b}
N. Linden and W. K. Wootters.
\newblock The Parts Determine the Whole in a Generic Pure Quantum State.
\newblock {\em Phys. Rev. Lett.} 89:277906, 2002.

\end{thebibliography}

\renewcommand{\thesection}{\Alph{section}}
\setcounter{section}{0}

\section{Equation and statement numbers in~\cite{lyonswalck}}

Table~1 gives a list of equation and statement numbers in the present
paper with their matching numbers in~\cite{lyonswalck}. 

\begin{table}[h]\label{statenumtable}
\begin{center}
  \begin{tabular}
    {|c|c|}\hline
number in this paper & number in~\cite{lyonswalck}\\ \hline
(\ref{akact}) & (10)\\
(\ref{bkact}) & (11)\\
(\ref{ckact}) & (12)\\
(4) & (13)\\
(5) & (1)\\
\ref{orbitdimasrank} & 3.3\\
(6) & (14)\\
\ref{triplesprop} & 5.1\\
\ref{mainpropgen} & 6.1\\
\ref{twocommon} & 5.3\\
\ref{twocommongen} & 6.2\\ 
\ref{minrankM} & 7.2\\
\ref{minorbthmstmnt} & 7.1\\ \hline
  \end{tabular}
\end{center}
\caption{Corresponding equation and statement numbers in~\cite{lyonswalck}}
\end{table}

\end{document}